# Position dependent photodetector from large area reduced graphene oxide thin films

Surajit Ghosh[1,*], Biddut K. Sarker[1,2], Anindarupa Chunder[1,3], Lei Zhai[1,3], and Saiful I. Khondaker[1,2,†]

[1]Nanoscience Technology Center, [2]Department of Physics, and [3]Department of Chemistry, University of Central Florida, Orlando, Florida 32826, USA.

We fabricated large area infrared photodetector devices from thin film of chemically reduced graphene oxide (RGO) sheets and studied their photoresponse as a function of laser position. We found that the photocurrent either increases, decreases or remain almost zero depending upon the position of the laser spot with respect to the electrodes. The position sensitive photoresponse is explained by Schottky barrier modulation at the RGO film-electrode interface. The time response of the photocurrent is dramatically slower than single sheet of graphene possibly due to disorder from the chemically synthesis and interconnecting sheets.

Due to its exceptional electrical and photonic properties, graphene is considered to be a promising material for optoelectronic devices [1-12]. Most importantly, graphene also has the ability to absorb photons over a wide spectrum from visible to infrared [5-12]. Recently there have been a number of interesting studies to investigate the photoresponse of this material. Park et al. used scanning phorocurrent microscopy to investigate photoresponse in single layer graphene sheet and found a strong photocurrent generation at the metal-graphene interface [5]. Xia et al. demonstrated ultrafast photodetectors using single layer and few layers graphene [6-7]. Lee et al. also used scanning photocurrent microscopy and reported local changes in the electronic structure of graphene sheets introduced by their interaction with deposited metal electrodes [8]. While many of these studies have been done on nanoscale devices based on single or few layers graphene, there is a growing interest in graphene based large area thin film devices for practical applications [9-15]. Thin films prepared from solution processed graphene nanostructures offer ease of material processing, low cost fabrication, mechanical flexibility, and compatibility with various substrates which make them an attractive candidate for large area devices. Graphene based thin films have already been used as a transparent and flexible material for electronic devices [13], active material in organic solar cell [10-12], as well as transparent electrodes in photovoltaic devices [14-15].

In this paper, we used graphene based thin films to fabricate large area position sensitive infrared photodetectors. The devices were fabricated by drop casting an aqueous suspension of chemically reduced graphene oxide (RGO) sheets and then making electrical contacts. We show that the photocurrent either increases, decreases or remain almost zero depending upon the position of the laser spot with respect to the electrodes. The maximum photo response [(light current-dark current)/dark current] was 193% when the laser was shined at electrode/film interface. The position sensitive photoresponse can be explained by the Schottky barrier modulation at the metal/RGO thin film interface. The time constant of the dynamic photoresponse was ~2.5 seconds which is much larger compared to the single sheets of graphene possibly due to disorder from the chemically synthesis and interconnecting sheets.

[*] Permanent Address: Department of Physics and Technophysics, Vidyasagar University, Midnapore 721 102, India
[†] To whom correspondence should be addressed. E-mail: saiful@mail.ucf.edu

RGO sheets were synthesized through a reduction of graphene oxide (GO) prepared by a modified Hummers method [16]. Oxidized graphite in water was ultrasonicated to achieve GO sheets followed by centrifugation for 30 minutes at 3000 rpm to remove any unexfoliated oxidized graphite. The pH of GO dispersion in water (0.1 mg/ml) was adjusted to 11 using a 5% ammonia aqueous solution. 15 µl of hydrazine solution (35% in DMF) was then added to the mixture. The mixture was heated at 95-100 $^0$C for 1 hour and cooled to room temperature. X-ray photoelectron spectroscopy (XPS) was taken before and after the reduction (figure 1(a)), shows that the intensity ratios of the C-C and C-O bonds changed from 1.4 to 8.0, verifying effective reduction of GO and are consistent with previous observations [17-18]. Subsequent atomic force microscopy (AFM) investigation (not shown here) of RGO suspension spin coated on a mica substrate shows 70% of the sheets are single layered with lateral dimension varying from 0.2 to 1.0 µm. The thin films were fabricated by drop casting the RGO suspension on a glass substrate. The solvent was allowed to evaporate in a fume hood for a few hours. The resulting films had an average thickness of ~ 0.6 µm. Figure 1(b) shows a scanning electrode micrograph (SEM) for one of our devices. The white regions in the SEM images indicate the edge or wrinkles of the RGO sheets. Finally, conducting silver paint was used to make pairs of parallel electrode of 32 mm separation.

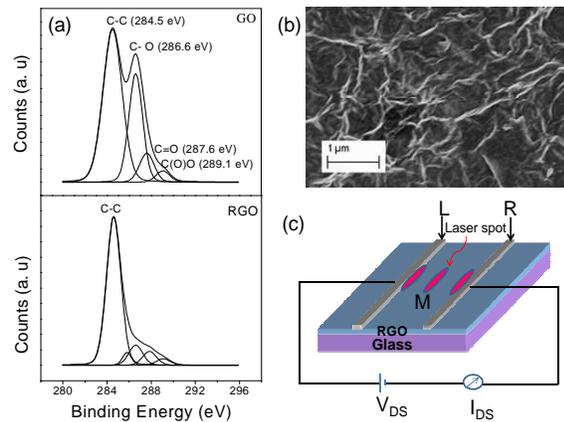

Figure 1. (color online) (a) X-ray photoelectron spectroscopy data of GO and RGO thin film on glass substrate (b) Scanning electron micrograph of RGO thin film. (c) Schematic diagram of the device and electron transport measurement setup. L, M and R are three different positions of NIR illumination.

Figure 1(c) shows a cartoon of our device and the electrical transport measurement setup. The dc charge transport measurement of the RGO films was carried out in a probe station in ambient conditions using a standard two probe method both in dark and under illumination by a near infrared (NIR) laser source positioned at three different locations. *L* corresponds to illumination on the left electrode/film interface, *R* corresponds to illumination on the right electrode/film interface and *M* corresponds to the middle position of the film. The NIR source consists of a semiconductor laser diode with peak wavelength of 808 nm and spot size of about 10 mm long by 1 mm wide. The diode laser was driven by a Keithley 2400 sourcemeter. The photointensity was measured using a calibrated silicon photo-diode (Thorlabs S121B). Unless mentioned otherwise, the intensity of the laser was 5.7 mW/mm$^2$ at the distance it was placed from the sample. Data were collected by means of LabView interfaced with the data acquisition card and current preamplifier (DL instruments: Model 1211) capable of measuring sub-pA signal. The photocurrent was calculated by subtracting the dark current from the current under NIR source illumination.

Figure 2(a) shows a typical photoresponse curve for one of our devices where we plot photocurrent as a function of time when the NIR source was illuminated at positions *L*, *M* and *R*. The photocurrent was measured under a small bias voltage $V_{bias}$ = 1 mV. The NIR source was turned on and off in 30 s intervals starting at t = 30 s. The plot is shown for two cycles of the laser being turned on and off to demonstrate the reproducibility of the data with time. The most interesting feature of this data is that the photocurrent strongly depends upon the position of the laser spot. At position *L* there is a large increase in photocurrent, while at position *R* there is a large decrease in photocurrent. Position *M* shows negligible change of photocurrent. The dark current of this device is 18.6 nA, while the current under illumination at position L is 54.5 nA giving an enhancement of 193%.

In order to determine the origin of the positional dependent photoresponse further, we measured the current-voltage (I-V) characteristics of the same device at *L*, *M*, and *R* under NIR illumination as well as in the dark condition, which is shown in figure 2(b).

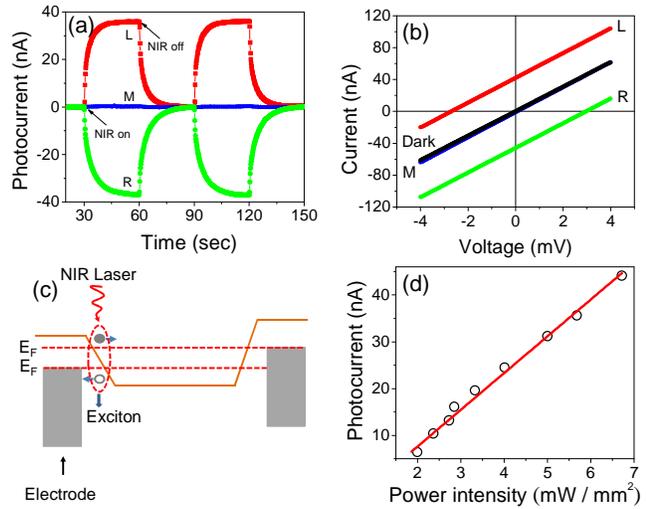

Figure 2. (color online) (a) Photocurrent as a function of time for a RGO film under NIR source illumination at positions L, M and R (b) Current-voltage characteristics of the device at three different illumination positions (L, R and M) and under dark condition (c) A cartoon for photovoltage generation at the metallic electrode / RGO interface. (d) Dependence of photo current with NIR source intensity.

The I-V curves for position *M* and in dark lie on top of one another and pass directly through the origin. Whereas, when illuminated at position *L* and *R*, the I-V curve is shifted above or below the origin respectively generating an offset voltage of about -2.7 and 3 mV respectively. A large enhancement of photocurrent as well as a finite photovoltage at the interface indicates that there is an existence of locally generated electric field at the metal-RGO film interface. This mechanism of the local electric field generation can be explained using Schottky barrier model schematically shown in figure 2(c) [6]. The black filled and open circles inside the dotted oval region represent the photogenerated exciton (bound electron-hole pair) due to absorption of NIR source (curved red arrow). Solid line is the potential variation within the graphene channel and dashed lines are the Fermi levels ($E_F$) of the two electrodes. When NIR is illuminated on the left electrode/RGO film interface, excitons are generated and dissociates into free charge carriers at the interface. Some of the free carriers (hole) might have sufficient energy to overcome the Schottky barrier and enter into the metal electrode leaving the electron in the film. This causes a hole-electron separation at the interface creating a positive photovoltage. Similar phenomenon occurs at the right electrode except that the right contact is a mirror image of the left contact and therefore the polarity of photovoltage and photoresponse reverses. However, at position M, the net photovoltage is zero since equal and opposite amount of electric field is generated at the two interfaces.

We now examine if it is possible that a thermal effect (bolometric) is also contributing to the photoresponse. From figure 2(b) we see that all the I-V curves are linear within the measured bias voltage of -4 mV to 4 mV. We calculate the resistance from the slope of the I-V curves at the different laser positions and in the dark. We find that the resistance holds a constant value of ~65 kΩ in the dark and at position L, M and R. If thermal effect was responsible, we would have expected a change in the resistance value upon laser illumination due to heating of the film. Additionally, we have also measured the temperature dependent I-V characteristics of one of our devices from room temperature to $200^0$C. We found that all I-V curves pass through origin (not shown here) without generating any offset voltage and the resistance decreases with increasing temperature. This is in contrast to what we have observed when NIR was illuminated. Therefore, we rule out any thermal effect in our photoresponse study.

We have also measured the photocurrent at different intensities of the NIR laser source at position L, shown in figure 2(d). The intensity of the laser was varied by changing the height between the sample and laser source. The open circles are the experimental data points and the solid line is a linear fit of the data showing the photo current increases linearly with intensity. When the intensity of the laser light is higher, more photons are absorbed by the RGO film and generates more excitons. So a greater number of holes have the probability to overcome the Schottky barrier, generating a larger photovoltage. On the other hand, when the intensity of laser light is low, a smaller photovoltage is generated.

We now study the time response of the photocurrent of the RGO film when NIR source was switched on and off. Figure 3(a) region I (t=0 to t=90 s) shows current versus time plot for one of the samples when the NIR laser was turned on at t = 30 second, held on for another 30 s and turned off at t = 60 s with $V_{bias}$ = 1.0 mV. It can be seen that when illuminated by the NIR source, the current increases slowly until it reaches a steady state and slowly recovers the dark current when the laser is switched off. In order to examine whether the slow time response has indeed come from the NIR illumination and not a delay due to RC like circuit existing in the entire setup we measured the response of the same sample with a step function voltage. This is shown in figure 3(a) region II (t=90 s to t=150 s). The bias was increased from 1.0 to 3.76 mV at t = 90 second, held on for the next 30 seconds, and then back to 1.0 mV at t=120 s. We found that unlike NIR source, the current increased almost instantaneously to a bias voltage increase. This verifies that the slow time response in region I is due to the NIR illumination. Interestingly, the voltage difference (2.76 mV) required to match the current due to the NIR is almost equal to that of the offset voltage (2.7 mV) determined from figure 2(b) (position L). The dynamic response of our device to the NIR source can be well described by $I(t) = I_{dark} + D\{1- \exp -(t-t_0)/\tau\}$ and $I(t) = I_{dark} + D \exp -(t-t_0)/\tau$ respectively for growth and decay respectively, where $\tau$ is the time constant, and $t_0$ is the time when NIR was switched on or off, $I_{dark}$ is the dark current and D is the scaling constant. This is shown in figure 3(b) (growth) and figure 3(c) (decay). The open circles are the experimental data points and solid lines are a fit to the above equations. From these fits, the time constant was calculated to be about 2.36 and 2.54 seconds for growth and decay respectively. The time response in our RGO film is much slower compared to what has been reported for individual graphene which is typically on the order of picoseconds [6]. In individual graphene sheet the time response can be ultrafast as the carrier transport is ballistic. However in the RGO film, the transport is dominated by disorder of the individual sheet as well as interconnection between different sheets leading to the diffusion of charge carriers, hence a slow time response.

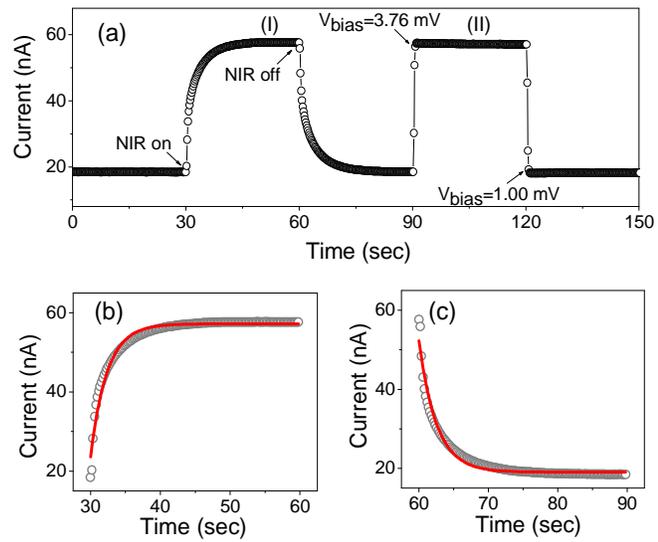

Figure 3. (color online) (a) Current as a function of time when NIR is illuminated at L (region I) and (II) with a bias voltage spike in dark (region II). Time response of photocurrent for (b) growth and (c) decay. The open circles are the experimental points and the red solid lines are fits to the exponential equations giving a time constant of 2.36 s and 2.54 s for growth and decay respectively.

In summary, we presented NIR photoresponse study of RGO thin films. We found that upon NIR laser illumination, the photocurrent increases, decreases or remain almost unchanged based on the position of the illumination. The maximum photoresponse of 193% was obtained at the metal electrode/RGO film interface is illuminated. The time response of the film is slow with a time constant of ~2.5 s. We explained the position dependent photoresponse by the Schottky barrier modulation at the metal/RGO thin film interface. This work show promise for the fabrication of large area low cost NIR photo detectors and position sensitive detectors using graphene nanostructures.

**Acknowledgments**
This work is partially supported by US National Science Foundation under grant ECCS 0801924. S.G. thanks DST, Government of India, for BOYSCAST fellowship.